# A remark on the negative activation volume for defects in solids

Vassiliki Katsika-Tsigourakou[1]

*Section of Solid State Physics, Department of Physics, National and Kapodistrian University of Athens, Panepistimiopolis, 157 84 Zografos, Greece*

ABSTRACT

The evolution of the aging process of glassy materials quenched from temperatures above their glass transition temperature $T_g$, when plotting the relaxed enthalpy versus the decrease in volume, leads to a slope comparable to the isothermal compressibility close to $T_g$. This empirical result was explained in an earlier publication (V. Katsika-Tsigourakou, G. E. Zardas J. Non-Cryst. Solids 356 (2010) 179-180) by means of a thermodynamical model. Here, we show that the same model enables the explanation of the rare cases of negative defect activation in solids.

*Keywords*: Dielectric properties, Relaxation, Glass transition, Defects

---

[1] Emai address: vkatsik@phys.uoa.gr

## 1. Introduction

In a previous publication [1] we studied the relation between relaxed enthalpy and volume during physical aging. In particular, time-depended phenomena referred to as physical aging are usually associated with the behavior of amorphous polymers or other glass-formers quenched from temperatures above their glass transition temperature $T_g$. An example of physical aging phenomenon is just the following: Upon stopping the cooling procedure at a temperature below $T_g$ and keeping this temperature constant, we measure a gradual decrease in volume $\upsilon$. This is accompanied by an amount of heat leaving the sample after the cooling procedure, which is the relaxed enthalpy $h$. Several workers [2-6] have studied the relaxed enthalpy versus volume plots, from which the following important results emerged: Upon comparing the slopes $dh/d\upsilon$ with measured compressibility ($\kappa$) data, a similarity between $dh/d\upsilon$ and $1/\kappa$ has been noticed [2, 3]. An explanation of this important result has been achieved in Ref. [1] on the basis of an early thermodynamical model [7, 8], which has been recently reviewed [9] and is shortly summarized below. It is the objective of this paper to show that the same thermodynamical model can explain experimental results, which were crossed with the publication of Ref. [1] and reported that negative defect activation volumes exist in.

## 2. The thermodynamical model

This model interconnects the point defect parameters in crystalline solids with the bulk elastic and expansivity properties. It suggests that the defect Gibbs energy $g^i$ is interconnected with the isothermal bulk modulus $B(=1/\kappa)$ through the relation $g^i = c^i B \Omega$ where $\Omega$ is the mean volume per atom and $c^i$ is a constant practically independent of temperature ($T$) and pressure ($P$). The superscript $i$ stands for the defect process studied, e.g., formation, migration, activation.

In Ref. [1] the following relation was derived (see Eq. (5) in Ref. [1]):

$$\frac{h^i}{\upsilon^i} = \frac{B - T\beta B - T\left(dB/dT\right)_P}{\left(dB/dP\right)_T - 1} \tag{1}$$

Where $\beta$ is the thermal volume expansion coefficient, $\upsilon^i = (dg^i/dP)_T$ stands for the defect volume and $h^i$ denotes the enthalpy for the corresponding defect process.

In Ref. [1] we argued that Equation (1) can be approximately written as:

$$\frac{\upsilon^i}{h^i} \approx \frac{1}{B}\left(\left.\frac{dB}{dP}\right|_T - 1\right) \tag{2}$$

(see Eq. (6) in Ref. [1]).

3. **The case of negative defect volumes**

Usually the quantity $\left.\frac{dB}{dP}\right|_T$ lies between 3 and 7, but there are rare cases in which it can be comparable or smaller than unity. Hence, Eq. (2) suggests that whenever $\left.\frac{dB}{dP}\right|_T < 1$ the quantity $\upsilon^i$ may be negative [10]. This can be alternatively realized as follows: Slater [11] using the Debye (D) approximation in a monoatomic solid, obtained the following relation for the (mean) Grüneisen constant γ:

$$\gamma = \frac{1}{2}\frac{dB}{dP}|_T - \frac{1}{6}$$

which was later improved in Refs [12, 13] to:

$$\gamma = \frac{1}{2}\frac{dB}{dP}|_T - \frac{1}{2} \tag{3}$$

from which one estimates the quantity $\frac{dB}{dP}|_T$ in terms of the Grüneisen constant:

$$\frac{dB}{dP}|_T = 2\gamma + 1 \tag{4}$$

By inserting Eq. (4) into Eq.. (2) we finally obtain:

$$\upsilon^i = 2h^i\gamma/B \tag{5}$$

respectively. Interestingly, Eq. (5) coincides with one derived by Flynn [14], within the Debye approximation in monoatomic crystals, on the basis of the dynamical theory for the defect migration in solids.

Recalling that Eq. (5) has been derived in the frame of the Debye approximation, we now seek for an appropriate relation in ionic crystals. In these crystals it is intuitively expected that the migration process takes place via the long wave-length TO modes (rather than the LO ones) [15]. For ionic crystals with solely central forces, the following expression for the frequency $\omega_{TO}$ of the transversal optical mode of zero wave number holds:

$$\omega_{TO} = \text{constant} \times (Bl/\mu)^{1/2} \tag{6}$$

where $l$ is the nearest-neighbor distance and $\mu$ is the reduced mass. Introducing Eq. (6) into the definition of the transverse optical Grüneisen constant

$$\gamma_{TO} \equiv -\frac{d\ln\omega_{TO}}{d\ln V}\Big|_T$$

(where V stands for the volume), we find:

$$\gamma_{TO} = \frac{1}{2}\frac{dB}{dP}\Big|_T - \frac{1}{6} \tag{7}$$

The values predicted from this relation have been found [17] to agree within 10% with the experimental values of a large number of ionic solids. Thus, relying on Eq. (7), we solve it in respect to $(dB/dP)|_T$ and upon inserting this result into Eq. (2) we find:

$$\upsilon^i = 2h^i(\gamma_{TO} - 1/3)/B \tag{8}$$

This relation reveals that $\upsilon^I$ becomes negative if $\gamma_{TO} < 1/3$.

## 4. Discussion and Conclusions

In other words, the above arguments lead to the conclusion that negative volumes for defect processes can be in principle detected. This is a paramount importance when considering relaxation processes of electric dipoles formed due to defects [16-18] arising from the presence of abiovalent impurities in ionic crystals. Negative defect volume means that the relaxation time $\tau$ of these electric dipoles may become smaller upon increasing pressure, thus leading to emission of transient electric signals before fracture. This is the generation mechanism [19] for the electric signals that are detected before major earthquakes [20-23]. This is understood in the context that before an earthquake the pressure (stress) gradually increases in the focal area before rupture. Actually recent laboratory measurements [18], the publication of which was crossed with that of Ref. [1], confirmed that negative activation volumes are observed in dielectric relaxation in hydrated rocks (see Table 1).

**References**

[1] V. Katsika-Tsigourakou, G.E. Zardas, J. Non-Cryst. Solids 356 (2010) 179.

[2] J. Hadac, P. Slobodian, P. Riha, P. Saha, R. W. Rychwalski, I. Emri and J. Kubat, J. Non-Cryst, Solids 353 (2007) 2681.

[3] P. Slobodian, P. Riha, R. W. Rychwalski, I. Emri, P. Saha, J. Kubat, Eur. Polym. J. 42 (2006) 2824.

[4] M. J. Kubat, J. Vernel, R. W. Rychwalski, J. Kubat, Polym. Eng. Sci. 38 (1998) 1261.

[5] P. Slobodian, P. Riha, A. Lengalova, J. Hadac, P. Saha, J. Kubat, J. Non-Cryst. Solids 344 (2004) 148.

[6] P. Slobodian, A. Lengalova, P. Saha, J. Therm. Anal. Cal. 71 (2003) 387.

[7] P. Varotsos, K. Alexopoulos, J. Phys. Chem. Solids 38 (1977) 997.

[8] P. Varotsos, Solid State Ionics 179 (2008) 438.

[9] P. Varotsos, K. Alexopoulos, Phys. Status Solidi B 110 (1982) 9.

[10] P. Varotsos, K. Alexopoulos, Phys. Rev. B 21 (1980) 4898.

[11] J. C. Slater, *Introduction to Chemical Physics* (New York: McGraw-Hill) 1939.

[12] J. S. Dugdale, D. K. C. Macdonald, Phys. Rev. 89 (1953) 832.

[13] Jai Shanker, A. P. Gupta, O. P. Sharma, Philos. Mag. B 37 (1978) 329.

[14] C. P. Flynn, *Point Defects and Diffusion* (Oxford: Clarendon) 1972.

[15] G. A. Samara, J. Phys. Chem. Solids 40 (1979) 509.

[16] M. Lazaridou, C. Varotsos, K. Alexopoulos, P. Varotsos, J. Physics C: Solid State 18 (1985) 3891.

[17] P. Varotsos, K. Alexopoulos, J. Phys. Chem. Solids 41 (1980) 443.

[18] P. Varotsos, K. Alexopoulos, J. Phys. Chem. Solids 42 (1981) 409.

[19] P. Varotsos, K . Alexopoulos, *Thermodynamics of Point Defects and their relation with the Bulk Properties* (Amsterdam: North Holland), 1986.

[20] P. Varotsos, K. Eftaxias, M. Lazaridou, K. Nomicos, N. Sarlis, N. Bogris, J. Makris, G. Antonopoulos, J. Kopanas, Acta Geophysica Polonica 44 (1996) 301.

[21] P. Varotsos, N. Sarlis, M. Lazaridou, Acta Geophysica Polonica 48 (2000) 141.

[22] P.A. Varotsos, N. V. Sarlis, E. S. Skordas, EPL (Europhysics Letters) 99 (2012) 59001.

[23] P.A. Varotsos, N.V. Sarlis, E.S. Skordas, Phys. Rev. E 66 (2002) 011902.

**Table 1.** Summary of the assumptions made here and the conditions found in order for the activation volume to be negative along with the results of recent laboratory measurements.

| No | Category of solids | Assumption made | Relation derived | Condition for negative $\upsilon^i$ |
|----|-------------------|-----------------|------------------|--------------------------------------|
| 1  | Any type          | $cB\Omega$ thermodynamical model | $\dfrac{\upsilon^i}{h^i} \approx \dfrac{1}{B}\left(\left.\dfrac{dB}{dP}\right|_T - 1\right)$ | $\left.\dfrac{dB}{dP}\right|_T < 1$ |
| 2  | Monoatomic        | Debye approximation | $\upsilon^i = 2h^i \gamma / B$ | $\gamma < 0$ |
| 3  | Ionic crystals    | Central forces  | $\upsilon^i = 2h^i(\gamma_{To} - \tfrac{1}{3})/B$ | $\gamma_{To} < 1/3$ |